# LUMINESCENCE OF a-QUARTZ

Anatoly Trukhin, Kaspars Truhins

Institute of Solid State Physics, University of Latvia, Kengaraga St.8, LV-1063 Riga, Latvia

**Abstract**

A short review of $\alpha$-quartz crystal's luminescence properties are presented. Among the host material's luminescence the luminescence of the self-trapped exciton (STE) is reviewed. This luminescence, which band is situated at 2.6-2.7 eV, could be observed mainly under ionising radiation with energetic yield about 20 %. The STE does not participate in pure recombination processes and could not be used in dosimetry. Host material defect luminescence at 5 eV appears in $\alpha$-quartz after heavy irradiation. It is constituted of permanent defect after neutron irradiation and transient defect after dens electron beam irradiation. This luminescence could be observed well at temperatures below 60 K. all another luminescence are of impurity nature. The Ge impurity luminescence in $\alpha$-quartz explained as STE near Ge. The aluminium and alkali complexes in $\alpha$-quartz provides at least three types of luminescence centers. One of them is with UV band at 6 eV, appears at low temperatures and could be excited only in tunnelling recombination process between pairs $[AlO_4 - Me^o]$, where $Me^o$ is an alkali ion captured an electron and a hole remains on aluminium tetrahedron. Another luminescence with band at 3.4 eV is also luminescence of complexes $[AlO_4/Me^+]$, which behaviour is similar to the luminescence of alkali alumo-silicate glass. The third luminescence with band at 3 eV could be observed mainly in natural $\alpha$-quartz, bright at temperatures below 200 K and is interpreted as STE-like luminescence at alumo-silicate clusters. The exchange of alkali ions to noble ions of copper of silver reduces original luminescence of alumo – alkali complexes and luminescence of noble ions appears. The main band of copper related luminescence is at 3.4 eV and that of silver is at 4.75 eV, both could be observed up to 500 K. their nature could be well described in terms of intra-ions transition. Exchange of noble ions back to alkali ions renews initial luminescence of the samples.

**Introduction**

Crystalline $\alpha$-quartz is widely used in luminescence dosimetry, however the picture of luminescence properties is not yet completed. Even known luminescence properties are spread among different paper related to describing of the nature of different luminescence in α-quartz. So, here we present short review of luminescence properties of the α-quartz.

As in all materials luminescence of α-quartz is of host material nature and impurity related. Among host material luminescence of α-quartz the luminescence of self-trapped excitons (STE) was studied in details and well-reasoned models are proposed (see for example review Trukhin, 2000). Less is known for host material defects' luminescence induced by ionizing radiation in α-quartz. Only neutron induced luminescence centers was discovered (Gee and Kastner, 1980) and that could be related to amorphous ranges of crystalline α-quartz , then informative for disordered state rather than to crystalline state properties.

Impurity related luminescence is studied in details for the case of germanium impurity, which creates a center of luminescence with properties very similar to that of STE. Therefore this center was ascribed to STE on Ge on tetrahedron. Also, the noble ions introduced into samples by exchange with alkali ions were studied . The model of centers in that case find good first explanation in the intra ion transition of the $Cu^+$ and $Ag^+$. The luminescence of significant technological impurities in α-quartz such as Al and alkali ions is very many-sided and in only some definite case there is some clearness in the nature of the luminescence. However models in these cases are not clear. In one case of natural quartz the similarity with STE in α-quartz were found and the center was ascribed to STE near alumo-alkali silicate crystalline claster, (Trukhin, Godmanis 1987).

**Self-trapped exciton luminescence**

Pure crystalline α-quartz exhibits very bright blue luminescence at temperatures lower than 180 K under ionizing irradiation with energetic yield about 25 %. It could not be excited in pure electron-holes recombination processes. For example, it could not be obtained in thermally stimulated luminescence. This luminescence was studied in many papers (Trukhin, Plaudis A., 1979; Griscom, 1979; Trukhin, 1982; Itoh, Tanimura, K. et al, 1989) and now there is consensus that it belongs to self-trapped excitons. The STE luminescence is of low quantum yield under photoexcitation in one-photon process and total photoluminescence of α-quartz is strongly affected by presence of impurities.

The main parameters of the STE luminescence are: two strongly overlapped luminescence broad bands with common maximum at 2.6-2.7 eV and FWHM about 1.4 eV, which total spectrum is shown in Fig.1. bands are resolved because of different energy of thermal quenching 0.1 and 0.2 eV and differently polarized with respect to crystal axes. The decay kinetics of both band is exponential with $\tau=1$ ms in the range of temperature 40 - 110K splitting to two components below 40 K becoming 0.33 ms and about 12 ms at 5-10 K, Fig.2. Slow kinetics as well as two components of decay at low temperatures both correspond to zero-field splitting (ZFS) of the triplet state of the STE. The fast component may represent transition to the ground singlet state from m=1or from m=-1, whereas slow component may represent to that from m=0. Such kinetics behaviour agrees to measured ODMR of the STE in α-quartz (Hayes et al, 1984) where the splitting parameters D and E of ZFS are determined. The D and E have unusually high value (D= 22 GHz and E = 1.5 GHz). The sequence on the model of STE from these data could be better understood in comparison with analogous decay kinetics data obtained in (Trukhin, 1993) for STE in $GeO_2$ crystal with α-quartz structure as well as with STE near germanium impurity in α-quartz, Fig.2 and see below.

These data are the base for model creation. Indeed, the efficiency of the spin-orbit interaction increases rapidly with atomic number, the effect of the interaction should be much larger in $GeO_2$, than in $SiO_2$ (McGlynn , et al, 1969). However, the decay time constant little changing with change of Si to Ge. The spin-orbit interaction allowing triplet-singler transitions matches to the states of oxygen. That opinion together with existence of the two kinds of STE, differing in luminescence polarization with respect to crystal orientation, and thermal activation energy of STE luminescence quenching, could be explained by STE model, proposed by (Trukhin, 1996), where the exciton self-trapping is started with electron appearance on antibonding state leading to Si-O bond weakening. Appeared in such way a non-bridging oxygen relaxes to direction of a bonding oxygen that the hole of STE was shared with this bonding oxygen. That provides fixation of STE with O-O bond creation of that NBO of STE with bonding oxygen on the other side of c or x,y channels, Fig.3. These two cases explain existence of the two kinds of STE with different energy of thermal quenching and different polarization of luminescence with respect of crystal orientation. The bond strength of quasi-molecule O-O determines energy of thermal quenching. This model should be valuable for crystals with α-quartz structure and that was proved for $SiO_2$ , $GeO_2$ , $AlPO_4$ , $GaPO_4$, (Trukhin, 1996).

**Self-trapped exciton luminescence on Ge impurity**

The α-quartz crystal with germanium impurity was studied by ESR (see, for example a review, Weil, 2000). The main conclusion is that germanium is substituting silicon in tetrahedron. Under ionizing radiation germanium traps electron and could create a series of centers in complexes with one valence ions (sodium, potassium, lithium and etc., Weil, 2000). It was found that germanium could trap a hole at sufficiently low temperatures (Hayes, Jenkin, 1985; Trukhin, 1986).

The luminescence of germanium in α-quartz was published in papers (Balodis, Valbis, 1973; Trukhin Plaudis, 1982; Trukhin 1987). The main parameters of the luminescence are very similar but not identical to the STE luminescence in pure quartz: two strongly overlapped luminescence broad bands with common maximum at 2.3 eV and FWHM about 1.4 eV (Fig.4) possessing different energy of thermal quenching 0.1 and 0.2 eV and different polarization with respect to crystal axes. The decay kinetics of both band is exponential with $\tau$=1.1 ms in the range of temperature 40 – 110 K and becomes split to two components below 40 K with time constants 0.36 ms and about 12 ms at 5-10 K, Fig.2, 5. The energetic yield (emitted energy per absorbed) is about 30% under x-ray irradiation (Trukhin, 1987). The quantum yield of the photoluminescence is also high with value about 0.4. That differs STE near Ge from host material STE.   The absorption threshold of pure α-quartz crystal is started at 8.5 eV where as that of Ge-doped samples is started at 7.4 eV.

**Luminescence related to presence of alkali ions**

Natural, as well as synthetic α-quartz crystal grows from alkali solution, therefore impurity of alkali ions is "technological". The alkali ions impurities could be detected by spectral analyses when burned, however the data on optical properties are not known well, with exception of correlation with radiation induced colour. The colour is explained on the base of ESR signal of hole trapped near aluminium, where an alkali ion was before irradiation as charge compensator (see, for example, Weil, 2000). Alkali ions could be exchanged on other one-valence ions such as proton, copper and silver ions (or analogous). The luminescence of alkali ions exhibits three different behaviours. Two of them are intra-center excited blue luminescence, one observed in natural samples, another observed in synthetic samples. Both also could be excited in recombination process. The third is UV luminescence, which could be excited only in recombination process. All kinds of alkali related luminescence could be removed by exchange to noble ions and renewed by reversal exchange of noble ions to some alkali ions ($Li^+$, $Na^+$, $K^+$).

**UV luminescence related to presence of alkali ions**

Natural (Brazil, Smoky) and synthetic crystalline quartz samples, containing large ($10^{18}$ cm$^{-3}$) concentration of aluminium impurity, as well as alkali ions as charge compensators, possess UV band. The last luminescence is situated in unusual UV range at 5.9-6 eV and was observed yet only in cathodoluminescence (CL), (Trukhin, Racko, Godmanis, 1989; Trukhin, Liblik, Lushchik, Jansons, 2003).

The luminescence intensity is significantly smaller (by about one order of magnitude) when the concentration of Al is on the level of $10^{17}$ cm$^{-3}$ and cannot be detected in samples with smaller Al content. Another CL band at 5 eV is observed in UV range of pure samples, (Trukhin, Liblik, Lushchik, Jansons, 2003) .

The intensity of the luminescence at 6 eV is completely disappeared in copper swept samples and renewed when those samples are swept with Li or Na ions. Then, the discovered luminescence is of impurity nature and directly connected with aluminium and alkali ions.

The UV luminescence decay kinetics under pulsed electron beam excitation is in microsecond time range, non-exponential with correspondence to power law $t^{-1}$ usually observed for tunnelling recombination processes. Such interpretation is in good agreement with the fact that decay kinetics is not depending on temperature in large interval, however luminescence intensity increases with decrease of the temperature. This luminescence is thermally quenched beyond 200 K.

Luminescence band is shifting during decay to high-energy side. This fact also corresponds to probable tunnelling nature of this luminescence. Therefore, in agreement with previous interpretation (Trukhin, Racko, Godmanis, 1989; Trukhin, Liblik, Lushchik, Jansons, 2003), the nature of luminescence is due to transitions between pairs Li$^o$-AlO$_4^+$ or Na$^o$-AlO$_4^+$, appeared under excitation. In the case of copper substituting alkali ions the transitions take place through the state of copper, which difference from alkali ions is in possibility to trap both electron and hole, whereas alkali ions, possessing very high second ionization potential, could not trap hole. It floats to the other hole state. Therefore we connect UV luminescence band at 6 eV with Al tetrahedrons having charge compensating alkali ions around.

**Luminescence of synthetic quartz**

In synthetic α-quartz beside UV luminescence there is another luminescence in blue-violet range of spectra correlating with presence of Al and of alkali ions. In such samples a very broad luminescence band at 3.4 eV excited at 7.7 eV takes place, Fig.6. This band as well as left side of excitation band is assigned "Alkali" in Fig.6. The excitation of this band starts from 7 eV. This luminescence becomes very low in intensity when the excitation photons' energies exceed 8.5 eV. This is the beginning of the fundamental absorption at 80 K. The non-exponential kinetics of luminescence of ms duration was observed (Fig.7). When samples are swept with noble ions (copper of silver) then luminescence changes it decay (Fig.7) and temperature dependence of intensity (Fig.8), therefore it is concluded that the initial luminescence belongs to removable ions and in our case these can be only alkali ions. It is necessary to underline that this initial luminescence of the synthetic quartz is very similar to the alkali ion luminescence in silica and alkali-silicate glasses (see, for example [12]). Such kind of luminescence is observable in relatively pure quartz, possibly without significant concentration of aluminium. Similarities with alkali silicate glasses could prove that this luminescence belongs to alkali ions dispersed in interstitial positions and luminescence is caused by creation of luminescence center oxygen-alkali ions. It is causes of differences between pure synthetic quartz from natural quartz.

**Violet luminescence related to presence of alkali ions in natural α-quartz**

It is generally accepted, that alkali ions are situated near aluminum substituting silicon in the tetrahedrons. Such point of view is proved by numerous investigations of radiation properties by ESR method (Weil, 2000), where hole trapping near aluminum is determined. Analogous situation is with germanium –alkali ions associations. Beside the signals are different for centers created at room temperature, when alkali ion could go away in thermally activated process, and at low (below 200 K), when alkali ion could not leave it position. However, it was discovered (Samoilovich, 1970; Trukhin, Godmanis, 1987) that proposed picture of structural position of Al and alkali ions is not complete. In the case of natural quartz partial contribution of center with Al related tetrahedron with alkali ions as a compensator of charge, could be low (about 10 – 20 %) (Samoilovich, 1970; Trukhin, Godmanis, 1987). Indeed, highest concentration of aluminium and corresponding alkali ions was detected in transparent part of some samples of smoky quartz with transparent core. In this part the highest concentration of luminescence center due to noble ions, $Cu^+$ and $Ag^+$ was obtained after exchange of alkali ions. Another part of this crystal possessing smoky colour gives only small amount of noble ions related center after exchange. It could not be attributed to single aluminium substituting silicon in a tetrahedron. So we could expect it belonging to some more complex of alumo-silicate clasters with charge compensators as alkali ions. This center could not be converted by irradiation into center responsible for smoky colour of quartz. Therefore a defect related to aluminum and alkali ions complex different from known defect responsible for smoky coloration. It structure is not yet known. This defect provide specific blue luminescence (Trukhin, Godmanis, 1987) called as low temperature luminescence, which spectral data are presented in Fig.9 and temperature dependences in Fig.10. This luminescence in natural quartz possesses the broad band at 3 eV and excitation for $h\nu > 5.8$ eV. Thermal quenching of this luminescence takes place for T>200K therefore it was started to call it as low temperature luminescence (LTL). The $\tau = 2$ ms at 80 K and $\tau_f$ is about 0.7 ms, $\tau_s = 12$ ms at 5 K, Fig.10. Quantum yield of LTL is high $\eta = 0.3$. Low temperature luminescence decay kinetics shows on a triplet-singlet transitions, because of long duration (~2 ms) and splitting into slow and fast component at LHeT. The last could match to zero field splitting of the triplet level. In spite of triplet nature of the low temperature luminescence and it correspondence to alkali ions content, there is no dependence of decay on the sort of alkali (Li, Na, K introduced in sample swept with copper ). Therefore there is no spin-orbital interaction with corresponding alkali ions, however they take part in center's structure. Similarities of LTL with behaviour of Ge in $\alpha$-quartz (Trukhin,1987) could attribute it to STE of alumo-silicate claster accompanied with alkali ions. Then some kind of oxygen molecule probably is created in that claster.

Most intensive this luminescence was obtained in annealed morions and was not detected in amethysts.

**$Cu^+$ and $Ag^+$ luminescence centers in $\alpha$-quartz**

After electrolysis of natural quartz sample in copper or argentums electrodes (Hetherington at al, 1965; Shendrik et al, 1973; Hohenau, 1984) a new luminescence bands of one-valence ions such as $Cu^+$ and $Ag^+$ appear. The initial luminescence disappeared. Two types of luminescence appear in both cases. One is low intensity band at low energy (2.5 eV for $Cu^+$ and 3.4 eV for $Ag^+$) and more intensive band at high energy (3.4 eV for $Cu^+$ and 4.75 eV for $Ag^+$), Fig.11,12. Significant, that in the case of silica glass after introducing of noble ions, the low energy band becomes principal, whereas high energy band is of low intensity. This correspondence allows us to ascribe low energy band center to a center where noble ion is connected to non-bridging oxygen. Non-bridging oxygen is one of the principal defect in silica glass, whereas in crystalline α-quartz it appears only in near surface area. The model for high energy band center is very depending on model of precursor defect. For the presented sample of natural quartz this defect gives low temperature luminescence, which disappears after introducing of the noble ions and reappears when noble ions are exchanged by alkali ions.

Copper and silver related luminescence could be well described by inner ion electronic transitions. The $Cu^+$ related PL excitation bands are starting from 4.5 eV to higher energy, Fig.11. PL intensity is almost independent on temperature in the range 5 - 300 K and is quenched at 500 K. The value of $\tau$ = 48 µs at temperature interval 20 - 350 K and at 5 K $\tau$ = 180 µs, Fig.11, insertion. The $Ag^+$ related center possesses several excitation bands starting from 5.3 eV, Fig.12 . Luminescence intensity is also independent on temperature up to 300 K and quenched at 500 K. The value of $\tau$ = 32 µs in temperature range 20 - 300 K and at 5 K $\tau$ = 80 µs, Fig.12, insertion. Increases of $\tau$ at low temperatures for both ions could be ascribed to triplet nature of excited states. Indeed, the value of $\tau$ is faster for silver than for copper and that could be related to increase of triplet-singlet transition probability in heavier ions due to spin-orbit interaction. The $Ag^+$ and $Cu^+$ are working as effective electron traps and after irradiation at 80 K many different $Ag^o$ and $Cu^o$ centers were observed in induced absorption and ESR spectra (Amanis et al, 1975; 1976), corresponding to atoms stabilization in many different interstitial places in lattice. At ambient temperature $Ag^o$ is moving and usually trapped by $Ag^+$ centers creating together $Ag_2^+$ centers or even $Ag_3^+$. Thus copper and silver introduced into α- quartz helps in understanding of alkali ions' behaviour under radiation.

Figures captions

Fig.1

The STE luminescence in quartz under photoexcitation can be observed only in extremely pure samples. In less pure samples the STE luminescence can be detected for excitation only in the range of 9 eV. There the level of STE luminescence yield is independent on sample purity. Insertion – excitation spectrum in wide range of energies.

Fig.2

The fast and slow components at low temperatures show evidence of the STE triplet state split in zero magnetic field. No trace of singlet-singlet luminescence was found. High temperature peculiarities show on two-stage thermal quenching of STE luminescence in quartz.

Fig.3

Excitation to antibonding state leads to Si-O bond weakening and the relaxation with an additional bond creation between NBO of STE and BO of opposite chain of $SiO_2$ tetrahedrons. The nature of this bond is partially due to covalent bonding and partially due to charge transfer (hole exchange among these NBO and BO). The O-O bond determines thermal stability of STE

Fig. 4.

Optical properties of $\alpha$-quartz crystal activated with germanium. PL-photoluminescence spectrum, PLE- photoluminescence excitation spectrum. The $\tau$ is luminescence decay time constant. Absorption spectrum of the Ge – doped samples is compared with that of pure sample. Measurements were performed on two samples with different Ge concentration.

Fig.5

Temperature dependences of luminescence parameters of $\alpha$-quartz crystal doped with Ge. I(T) is PL intensity, P(T) is polarization of PL, $\tau$(T) is decay kinetics time constant. $\tau_{low}$ is $\tau$ of centers' with low value of activation energy for thermal quenching (0.2 eV) and for $\tau_{high}$ it is (0.3 eV). $\tau_{fast}$ and $\tau_{slow}$ are constants for two components of decay corresponding to zero field splitting of the triplet level.

Fig.6

Photoluminescence emission and excitation spectra of synthetic $\alpha$-quartz of different purity: "Alkali" – an as received sample possessing photoluminescence excitable at 7.7 eV; "Cu" – an as-received sample activated with copper; "STE" – the self-trapped exciton spectre of the preceding two sample and a purer sample. Measurements were made of 80 K except for "Cu", which are measured at 290 K.

Fig.7

Photoluminescence decay kinetics curves in α-quartz samples, which spectra are presented in Fig.6. T=80 K.

Fig.8

Temperature dependence of the intensity of the luminescence in synthetic α-quartz by excitation at the peak of the bands in the excitation spectra; alkali-band at 8.3 eV due to alkali related centers, Cu band at 7.7 eV due to cu-related centers, and STE.

Fig. 9

Optical spectra of natural crystalline quartz possessing low temperature luminescence (the sample is smoky quartz with transparent core, for which photoluminescence intensity is 4 times higher than smoky faces). PL – photoluminescence; PLE – photoluminescence excitation.

Fig.10

Temperature dependences of PL intensity and it decay time constant for excitation with 6 eV photon of the natural α-quartz crystal sample, which spectra are presented in Fig.9.

Fig. 11

Photoluminescence (PL) and it excitation (PLE) spectra of natural crystalline α-quartz activated with copper by exchange of alkali ions to copper ions by high temperature (800 $^{o}$C) electrolyses of the sample (spectra of the sample before treatment are presented in the Fig.9). insertion – temperature dependence of the PL decay time constant (independent on excitation energy within PLE spectrum).

Fig.12

Photoluminescence (PL) and it excitation (PLE) spectra of natural crystalline α-quartz activated with silver by exchange of alkali ions to silver ions by high temperature (800 $^{o}$C) electrolyses of the sample (spectra of the sample before treatment are presented in the Fig.9). insertion – temperature dependence of the PL decay time constant (independent on excitation energy within PLE spectrum).

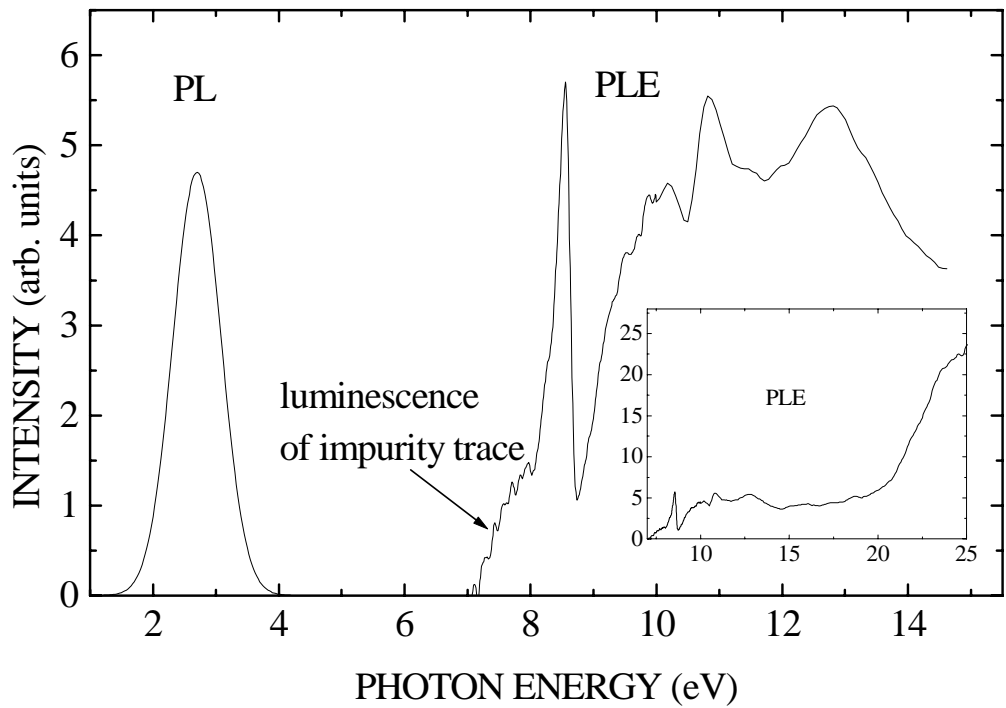

Fig.1

The STE luminescence in quartz under photoexcitation can be observed only in extremely pure samples. In less pure samples the STE luminescence can be detected for excitation only in the range of 9 eV. There the level of STE luminescence yield is independent on sample purity. Insertion – excitation spectrum in wide range of energies.

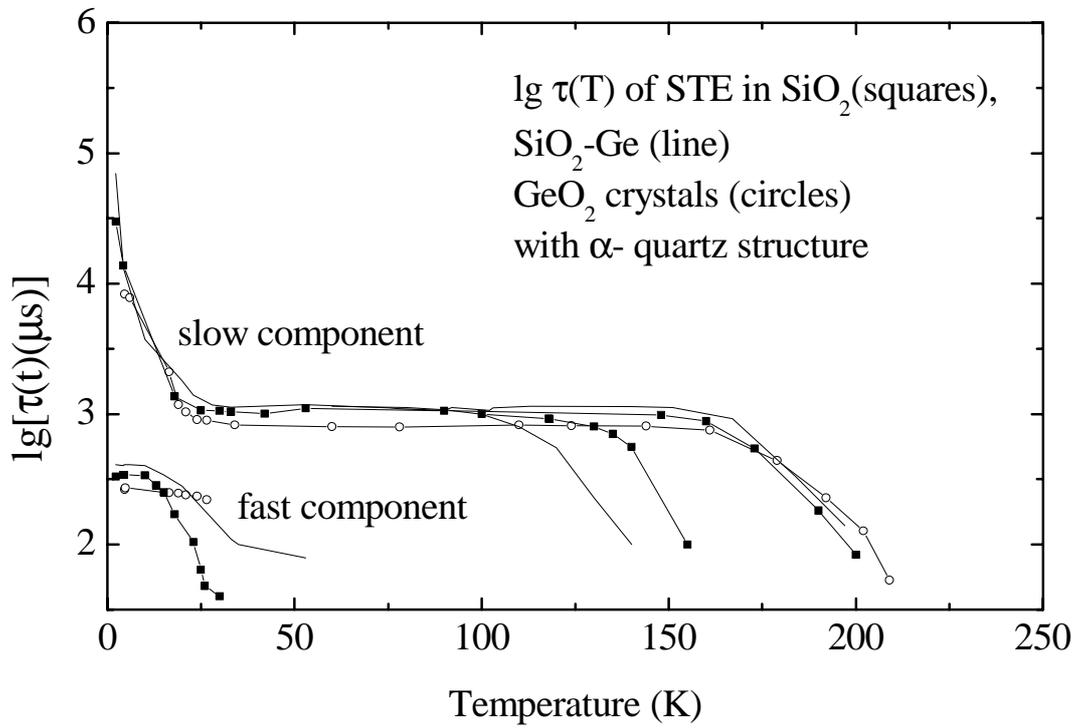

Fig.2

The fast and slow components at low temperatures show evidence of the STE triplet state split in zero magnetic field. No trace of singlet-singlet luminescence was found. High temperature peculiarities show on two-stage thermal quenching of STE luminescence in quartz.

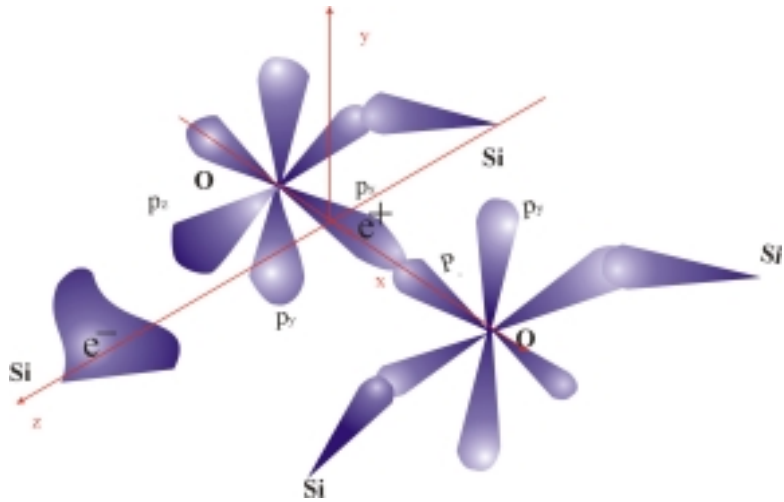

Fig.3

Excitation to antibonding state leads to Si-O bond weakening and the relaxation with an additional bond creation between NBO of STE and BO of opposite chain of $SiO_2$ tetrahedrons. The nature of this bond is partially due to covalent bonding and partially due to charge transfer (hole exchange among these NBO and BO). The O-O bond determines thermal stability of STE.

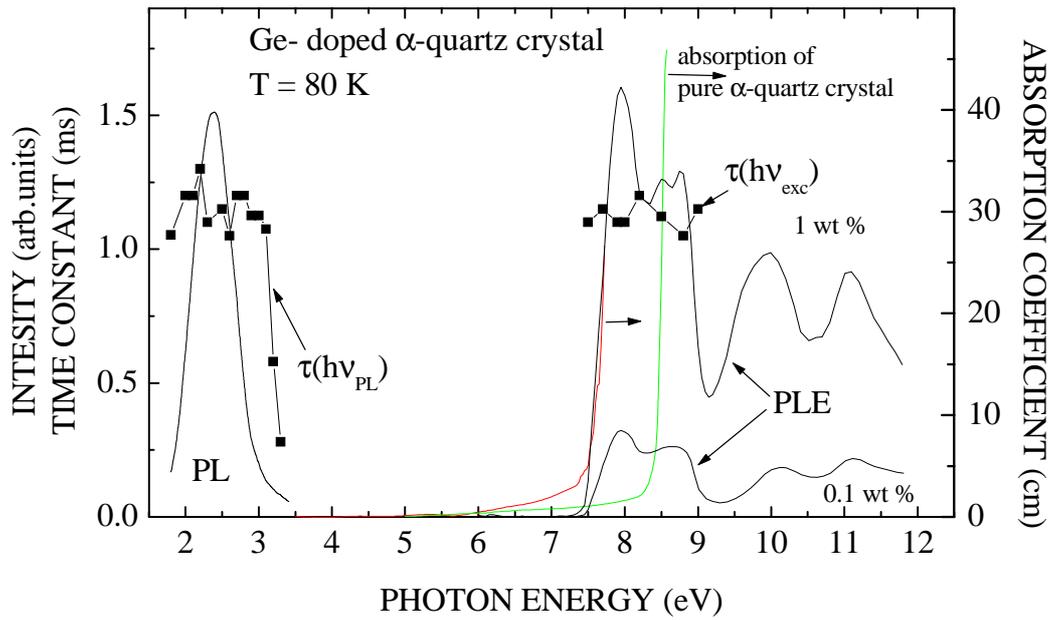

Fig. 4.

Optical properties of α-quartz crystal activated with germanium. PL-photoluminescence spectrum, PLE- photoluminescence excitation spectrum. The τ is luminescence decay time constant. Absorption spectrum of the Ge – doped samples is compared with that of pure sample. Measurements were performed on two samples with different Ge concentration.

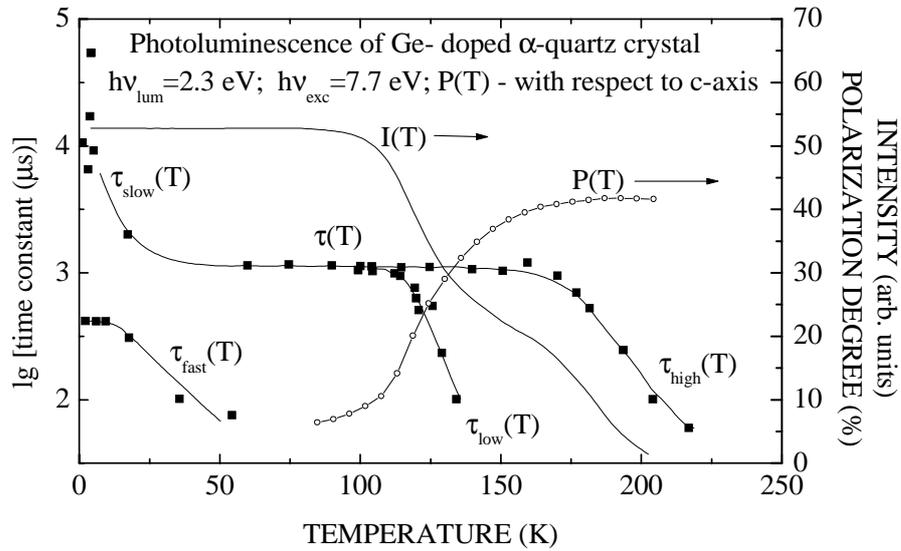

Fig.5

Temperature dependences of luminescence parameters of α-quartz crystal doped with Ge. I(T) is PL intensity, P(T) is polarization of PL, τ(T) is decay kinetics time constant. $\tau_{low}$ is τ of centers' with low value of activation energy for thermal quenching (0.2 eV) and for $\tau_{high}$ it is (0.3 eV). $\tau_{fast}$ and $\tau_{slow}$ are constants for two components of decay corresponding to zero field splitting of the triplet level.

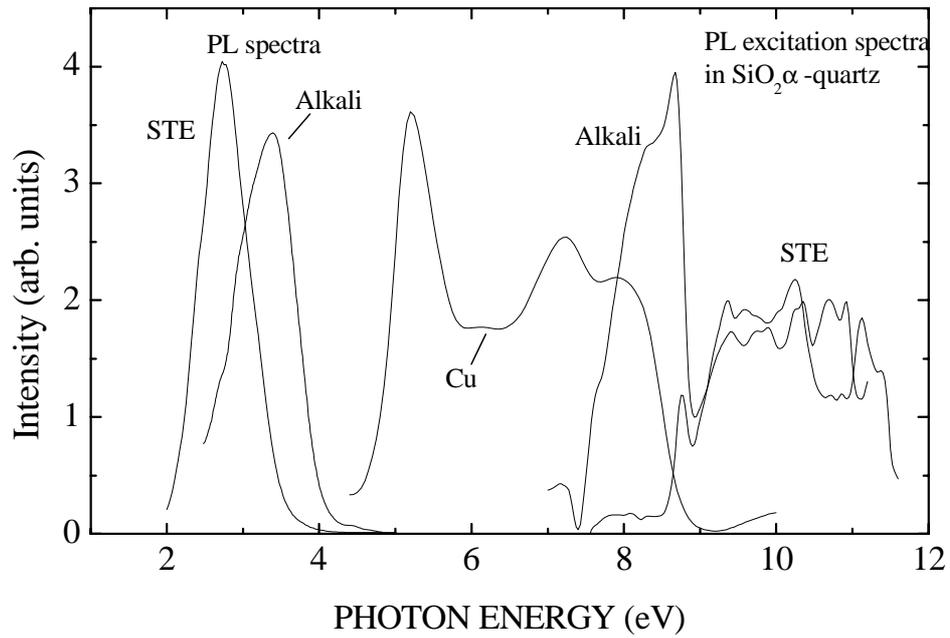

Fig.6

Photoluminescence emission and excitation spectra of synthetic α-quartz of different purity: "Alkali" – an as received sample possessing photoluminescence excitable at 7.7 eV; "Cu" – an as-received sample activated with copper; "STE" – the self-trapped exciton spectre of the preceding two sample and a purer sample. Measurements were made of 80 K except for "Cu", which are measured at 290 K.

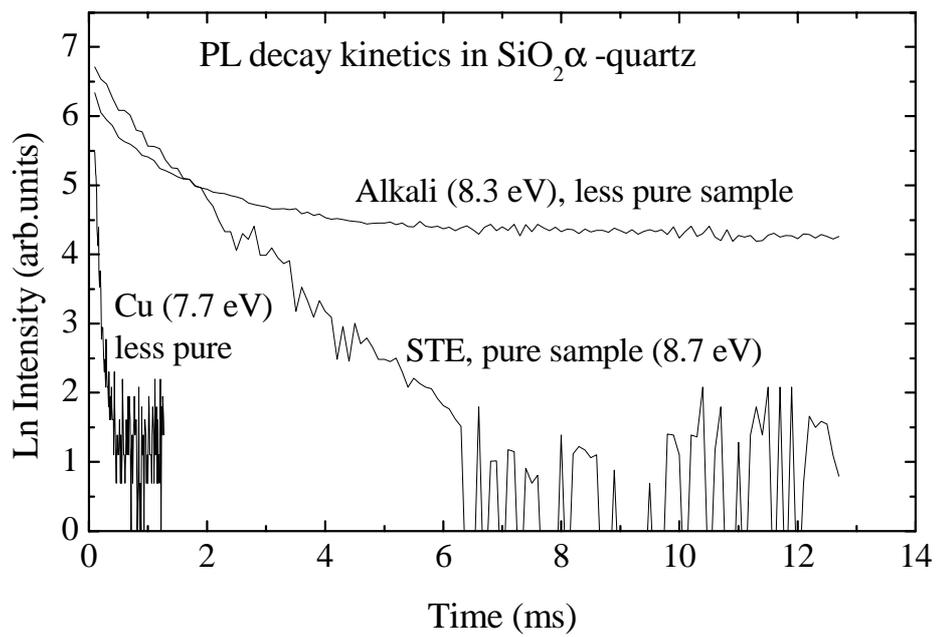

Fig.7

Photoluminescence decay kinetics curves in α-quartz samples, which spectra are presented in Fig.6. T=80 K.

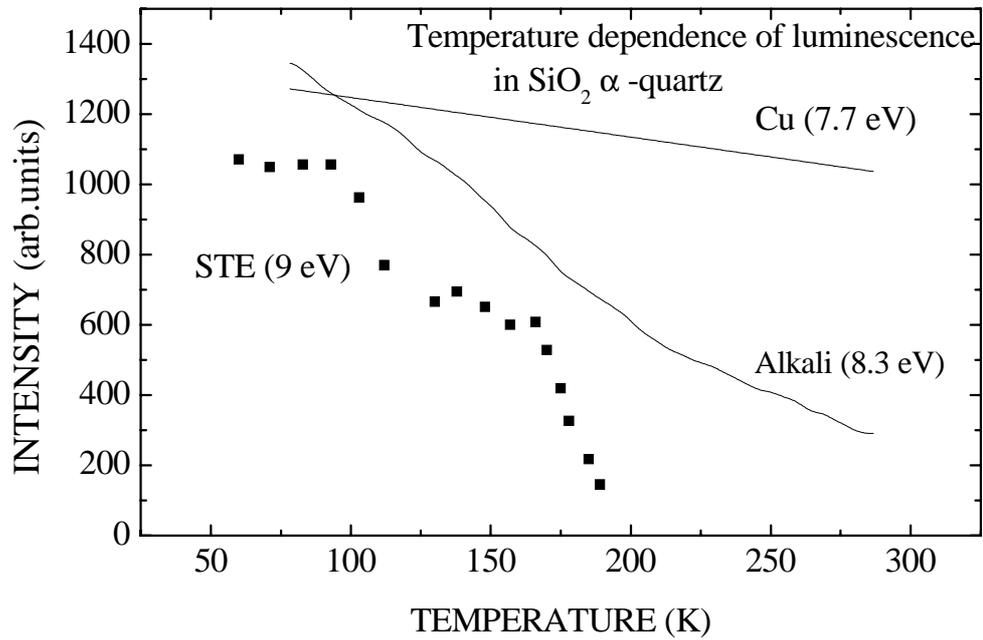

Fig.8

Temperature dependence of the intensity of the luminescence in synthetic α-quartz by excitation at the peak of the bands in the excitation spectra; alkali-band at 8.3 eV due to alkali related centers, Cu band at 7.7 eV due to cu-related centers, and STE.

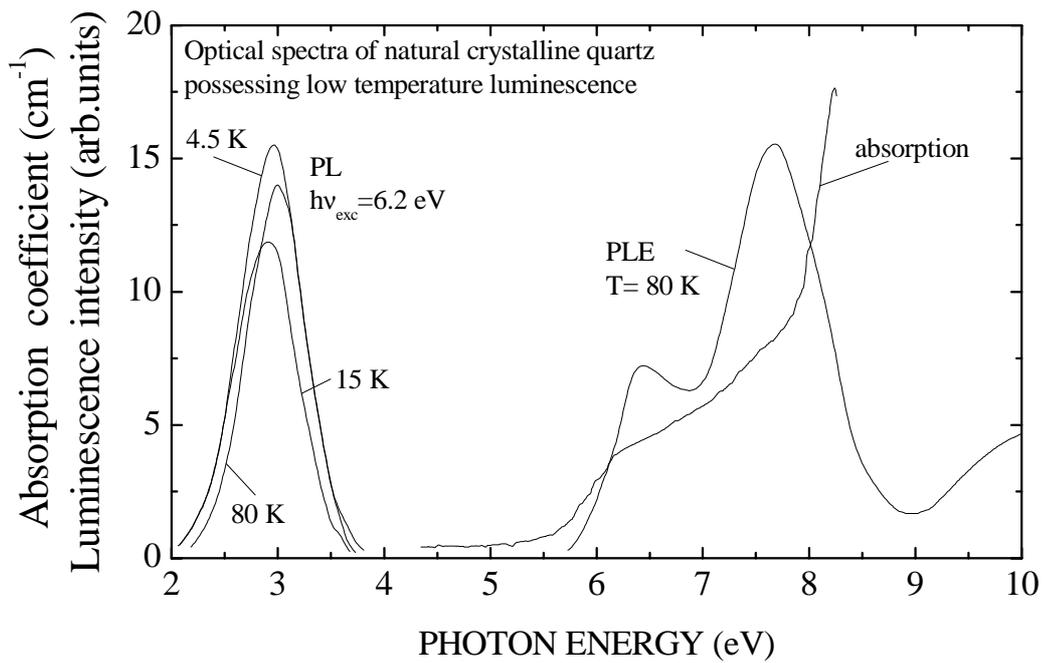

Fig. 9

Optical spectra of natural crystalline quartz possessing low temperature luminescence (the sample is smoky quartz with transparent core, for which photoluminescence intensity is 4 times higher than smoky faces). PL – photoluminescence; PLE – photoluminescence excitation.

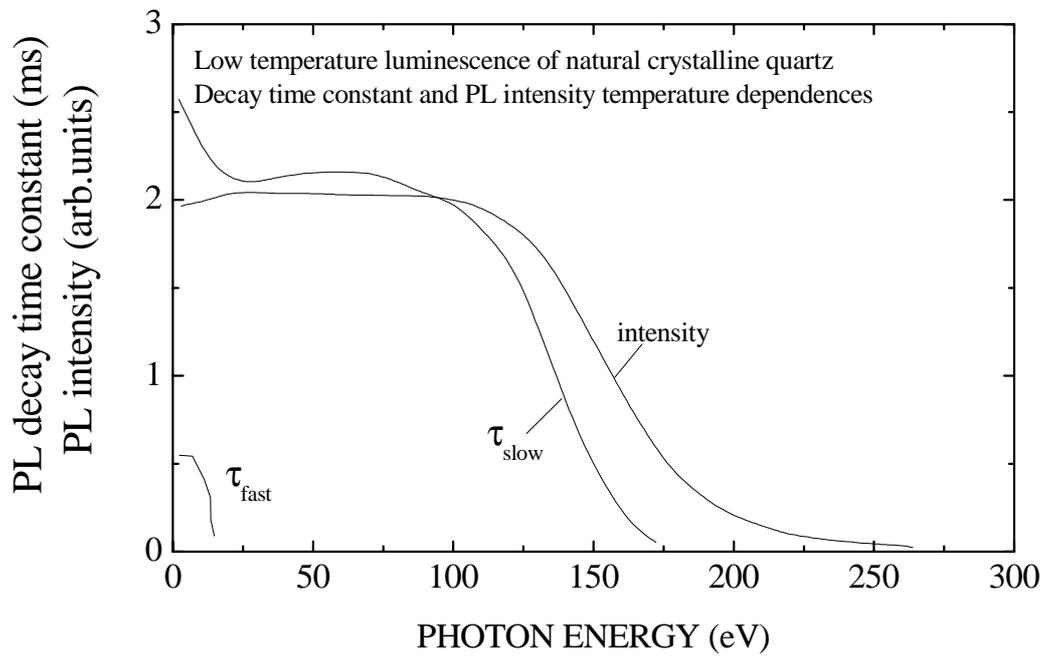

Fig.10

Temperature dependences of PL intensity and it decay time constant for excitation with 6 eV photon of the natural α-quartz crystal sample, which spectra are presented in Fig.9.

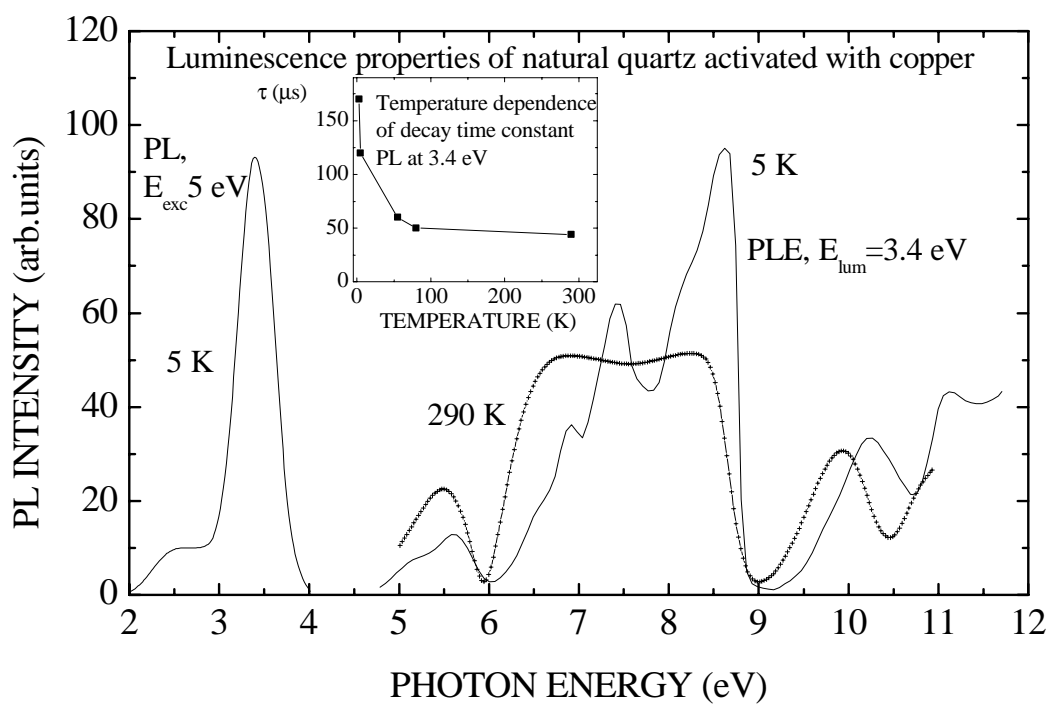

Fig. 11

Photoluminescence (PL) and it excitation (PLE) spectra of natural crystalline α-quartz activated with copper by exchange of alkali ions to copper ions by high temperature (800 °C) electrolyses of the sample (spectra of the sample before treatment are presented in the Fig.9). insertion – temperature dependence of the PL decay time constant (independent on excitation energy within PLE spectrum).

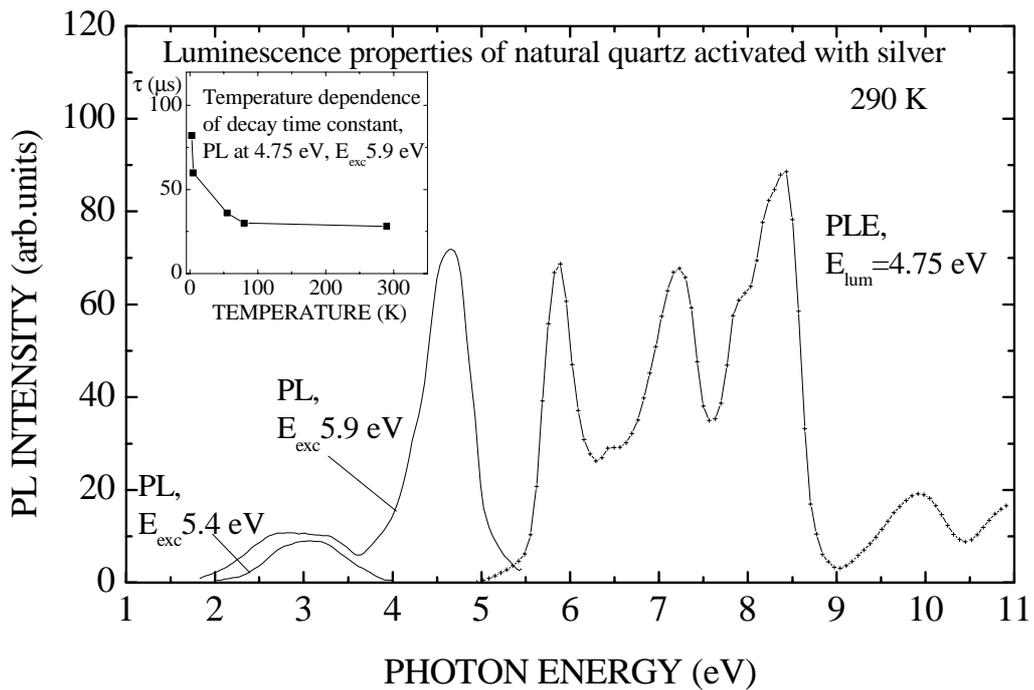

Fig.12

Photoluminescence (PL) and it excitation (PLE) spectra of natural crystalline α-quartz activated with silver by exchange of alkali ions to silver ions by high temperature (800 °C) electrolyses of the sample (spectra of the sample before treatment are presented in the Fig.9). insertion – temperature dependence of the PL decay time constant (independent on excitation energy within PLE spectrum).